\documentclass[showpacs,preprintnumbers,amsmath,amssymb]{revtex4}

\usepackage{graphicx}
\usepackage{dcolumn}
\usepackage{bm}
\usepackage{mathrsfs}
\usepackage{color}
\usepackage{comment}

\newcommand{\bvec}[1]{\mbox{\boldmath $#1$}}

\newcommand{\ket}[1]{\left|#1\right\rangle}

\begin{document}
\title{Effects of cluster-shell competition and BCS-like pairing in $^{12}\mathrm{C}$}

\author{H.~Matsuno${}^1$ and N.~Itagaki${}^2$}


\affiliation{$^1$Department of Physics, Kyoto University, Kitashirakawa Oiwake-Cho, Kyoto 606-8502, Japan\\
$^2$Yukawa Institute for Theoretical Physics, Kyoto University, Kitashirakawa Oiwake-Cho, Kyoto 606-8502, Japan}


\begin{abstract}
The antisymmetrized quasi-cluster model (AQCM) was proposed to describe $\alpha$-cluster and $jj$-coupling shell models on the same footing. 
In this model, the cluster-shell transition is characterized by two parameters; $R$ representing the distance between $\alpha$ clusters and $\varLambda$ describing the breaking of $\alpha$ clusters, and the contribution of the spin-orbit interaction, very important in the $jj$-coupling shell model, can be taken into account starting with the $\alpha$ cluster model wave function. 
Not only the closure configurations of the major shells, but also the subclosure configurations of the $jj$-coupling shell model can be described starting with the $\alpha$-cluster model wave functions; however, the particle hole excitations of single particles have not been fully established yet.
In this study we show that the framework of AQCM can be extended even to the states with the character of  single particle excitations.
For $^{12}\mathrm{C}$, two particle two hole (2p2h) excitations from the subclosure configuration of $0p_{3/2}$ corresponding to BCS-like pairing are described, and these shell model states are coupled with the three $\alpha$ cluster model wave functions.
The correlation energy from the optimal configuration can be estimated not only in the cluster part but also in the shell model part.
We try to pave the way to establish a generalized description of the nuclear structure.
\end{abstract}



\maketitle

\section{Introduction}
One of the important goals of nuclear structure physics is the description of both shell and cluster aspects on the same footing.
The real nuclear systems have both characters, and the mixing, or competition of these two, is an important subject for the physics of quantum many-body systems~\cite{CSC,Suhara.PhysRevC87.054334.2013,Suhara2015}.
Our strategy is to establish a framework, which starts with the cluster model side, contrary to standard approaches, and includes shell correlations. \par
As it is well-known, when we take zero limit for the relative distances between clusters, the model space coincides with that of the lowest shell model configuration.
This is called SU(3) limit~\cite{elli58}, and for $N=Z$ nuclei with magic numbers of three dimensional harmonic oscillator ($N=Z=2,8,20,\ldots$), cluster model wave functions agree with the doubly closed shell configurations.
Here both spin-orbit favored ($j$-upper) and unfavored ($j$-lower) single particle orbits are filled and we can forget about the spin-orbit contribution.
However, the spin-orbit effect exists in other cases, and in most of the conventional cluster models, this effect cannot be taken into account; the spin-orbit contribution cancels because of the assumption of $\alpha$ cluster that four nucleons have the same form of the spatial wave function. \par
To overcome this difficulty of the cluster model, we proposed antisymmetrized quasi-cluster model (AQCM)~\cite{Itagaki.PhysRevC71.064307,Masui,Yoshida2,Ne-Mg,Suhara.PhysRevC87.054334.2013,Suhara2015,Itagaki2016,Matsuno2017}, which enables us to describe the $jj$-coupling shell model states with the spin-orbit contribution starting with the cluster model wave function.
In AQCM, the transition from the cluster- to shell-model-structure can be described by two parameters; $R$ representing the distance between $\alpha$ clusters, and $\varLambda$, which characterizes the transition of $\alpha$ cluster(s) to quasi-cluster(s) and quantifies the role of the spin-orbit interaction.
In Ref.~\cite{Suhara.PhysRevC87.054334.2013}, the AQCM wave function was shown to correspond to the $(0s_{1/2})^4 (0p_{3/2})^8$ closed shell configuration of $^{12}\mathrm{C}$, and strong contribution of the spin-orbit interaction was taken into account. 
The optimal ground state of $^{12}\mathrm{C}$ was shown to have an intermediate character between the three $\alpha$ clusters and shell model states.
In a similar way, the subclosure configuration of $0d_{5/2}$ was described in $^{28}\mathrm{Si}$, and characteristic magic numbers of the $jj$-coupling shell model, $28$ and $50$, were successfully described in $^{56}\mathrm{Ni}$ and $^{100}\mathrm{Sn}$~\cite{Itagaki2016}.
\par
However, the particle hole excitations of single particles are not fully established yet from cluster model point of view.
The purpose of the present study is to show that the framework of AQCM can be extended even to the states with the character of single particle excitations. 
The first example is $^{12}\mathrm{C}$.
Some configurations, which are excited from the subclosure configuration of $0p_{3/2}$ of the $jj$-coupling shell model, are introduced, and the effects of BCS-like pairing for the proton part, neutron part, and proton-neutron part are taken into account. 
Also the coupling effect with the cluster states is investigated. \par
So far the features of $^{12}\mathrm{C}$ have been investigated using many different models; various cluster models~\cite{Kamimura.NuclPhysA351.456.1981,Uegaki.ProgTheorPhys57.1262.1977,Funaki.PhysRevC67.051306R.2003}, 
shell models including modern \textit{ab initio} ones~\cite{Cohen-Kurath,Navratil.PhysRevC68.034305.2003}, and so on.
The $0_2^+$ state, which is known as the Hoyle state, is nicely described by the three $\alpha$ cluster models; however they cannot describe detailed properties related to the $\alpha$ cluster breaking effect especially in the ground state rotational band.
On the other hand, in principle the shell model provides a complete set, but the cluster states are in practice difficult to be described within finite model space.
Takigawa \textit{et al.} have introduced a hybrid model to mix $\alpha$ cluster model and $p$ shell SU(3) basis states~\cite{Takigawa}.
Our spirit is based on this idea; however we transform the cluster model wave functions directly to the ones of the $jj$-coupling shell model and try to pave the way to establish a generalized description of the nuclear structure.
Also, antisymmetrized molecular dynamics (AMD) and Fermionic molecular dynamics (FMD) have been successfully introduced to describe both characters of shell and cluster models~\cite{AMD-1,AMD-2,FMD-1,FMD-2,FMD-3}.
In these models, central positions of all the nucleons are optimized under some constrains.
On the other hand, in our approach, we introduce much fewer and controllable parameters, which allow the description of excited configurations.
\par
In AQCM, we transform Brink-type $\alpha$ cluster model wave function~\cite{Brink} to the $jj$-coupling shell model wave function by giving imaginary part for the Gaussian center parameters.
This procedure has some similarity with the idea of Fock-Bargmann space developed by Filippov \textit{et al.}~\cite{Filippov}.
In Ref.~\cite{Filippov}, they discussed $^6\mathrm{He}$ and the hyperspherical harmonics basis states have been introduced for the description of two valence neutrons outside of the $\alpha$ core, and the matrix elements of the Hamiltonian have been extracted from the expectation value obtained by using a Gaussian wave packet.
We also use Gaussian wave packets; however, in our study, we directly transform the wave function to the $jj$-coupling shell model and the breaking effect of the $\alpha$ cluster part can be discussed. \par
This paper is organized as follows.
We describe our formulation in this work including the review for AQCM in Sec.~\ref{Sec.Formulation}.
The results and discussion are given in Sec.~\ref{Sec.ResultsAndDiscussion}.
Finally, we present conclusion and outlook in Sec.~\ref{Sec.Conclusion}.

\section{Formulation}
\label{Sec.Formulation}

\subsection{AQCM wave function}
As in many conventional models, the single-particle wave function of AQCM ($\phi_i$) consists of the spatial ($\psi_i$), spin ($\chi_i$), and isospin ($\tau_i$) parts,
\begin{align}
\phi_i=\psi_i\chi_i\tau_i.
\label{Eq.SingleParticleWaveFunction}
\end{align}
The spatial part of the single-particle wave function has a Gaussian shape~\cite{Brink},
\begin{align}
\psi_i=\left(\frac{2\nu}{\pi}\right)^{\frac{3}{4}}\exp[-\nu(\bvec{r}-\bvec{\zeta}_i)^2],
\label{Eq.SpatialPart}
\end{align}
where $\nu$ is the width parameter.
From these single-particle wave functions, the Slater determinant of $A$ nucleon system $\varPsi=\mathcal{A}[\phi_1,\ldots,\phi_A]$ is constructed, where $\mathcal{A}$ is the antisymmetrizer for all nucleons.
If we give the same value for the Gaussian center parameter $\bvec{\zeta}_i$ of four nucleons (spin-up proton, spin-down proton, spin-up neutron, and spin-down neutron) as in the so-called Brink model, they form an $\alpha$ cluster, and the contribution of the spin-orbit interaction vanishes because of the antisymmetrization effect. \par
In Ref.~\cite{Itagaki.PhysRevC71.064307}, the AQCM wave functions were shown to describe subclosure configurations of the $jj$-coupling shell model.
The Gaussian center parameters $\{ \bvec{\zeta}_i \}$ are complex vectors, and the imaginary parts are introduced as
\begin{align}
\mathrm{Im}\bvec{\zeta}_i=\varLambda\bvec{e}_i^{(\mathrm{spin})}\times\mathrm{Re}\bvec{\zeta}_i,
\label{idea}
\end{align}
where $\bvec{e}_i^{(\mathrm{spin})}$ is a unit vector for the intrinsic-spin orientation of $i$-th nucleon, and $\varLambda$ is an order parameter for the dissolution of the cluster.
By introducing $\varLambda$, $\alpha$ clusters are transformed to quasi clusters with the spin-orbit contribution.

\subsection{Description of subclosure configuration ($^{12}\mathrm{C}$ case)}
\label{Sec.DescriptionOfSubclosureConfiguration12CCase}
Before extending AQCM to describe single particle excitations, here we review the description of the subclosure configuration for the $^{12}\mathrm{C}$ case~\cite{Suhara.PhysRevC87.054334.2013}.
This part is the mathematical interpretation of AQCM and not needed in the actual calculation; however we have to recall the important parts for further extension of the model.
Since the neutron part is introduced in the completely same way, here we concentrate on the proton part.
The protons $i=1$ and $2$ are in a common quasi cluster with spin-up and spin-down.
Based on the original idea of Eq.~(\ref{idea}), the Gaussian center parameters are introduced as
\begin{align}
\bvec{\zeta}_{i=1}=R(\bvec{e}_x+i\varLambda\bvec{e}_y)
\label{Eq.zeta1}
\end{align}
and
\begin{align}
\bvec{\zeta}_{i=2}=R(\bvec{e}_x-i\varLambda\bvec{e}_y),
\label{Eq.zeta2}
\end{align}
where $\bvec{e}_x$ and $\bvec{e}_y$ are unit vectors on the $x$ and $y$ axes, respectively.
There are put on the $x$ axis, and imaginary parts are given in the $y$ and $-y$ directions, since their intrinsic spins are quantized along the $z$ axis ($z$ and $-z$ directions).
They are introduced as time reversal partners.
The squares in the powers of the single-particle wave functions can be expanded as
\begin{align}
\phi_{i=1}=\left(\frac{2\nu}{\pi}\right)^{\frac{3}{4}}\exp[-\nu(\bvec{r}^2+\bvec{\zeta}_1^2)+2\nu\bvec{r}\cdot\bvec{\zeta}_1]\chi_\uparrow\tau_1, 
\label{spwf1}
\end{align}
\begin{align}
\phi_{i=2}=\left(\frac{2\nu}{\pi}\right)^{\frac{3}{4}}\exp[-\nu(\bvec{r}^2+\bvec{\zeta}_2^2)+2\nu\bvec{r}\cdot\bvec{\zeta}_2]\chi_\downarrow\tau_2,
\end{align}
where $\chi_\uparrow$ and $\chi_\downarrow$ stand for spin-up and down, respectively, and $\tau_1$ and $\tau_2$ are isospin wave functions of the protons. 
In Eq.~(\ref{spwf1}), the cross-term part in the power of the exponential can be Taylor expanded, and by substituting Eq.~(\ref{Eq.zeta1}), this factor is described as
\begin{align}
\exp[2\nu\bvec{r}\cdot\bvec{\zeta}_1]=\sum_{l=0}^\infty\frac{1}{l!}(2\nu{}Rr)^l\left(\frac{x+i\varLambda{}y}{r}\right)^l.
\label{Taylor}
\end{align}
For $\varLambda=1$, by using the spherical harmonics $Y_{lm}(\varOmega)$ and introducing the radial part of the spatial wave function $R_{0l}(r)$, the single-particle wave function of the proton $i=1$ can be expressed as
\begin{align}
\phi_{i=1}=\left(\frac{2\nu}{\pi}\right)^{\frac{3}{4}}\sum_{l=0}^\infty\frac{(2\nu{}R)^l}{l!s_lt_l}R_{0l}(r)Y_{ll}(\varOmega)\chi_\uparrow\tau_1,
\end{align}
where 
\begin{align}
\left(\frac{x+iy}{r}\right)^l=\frac{1}{s_l}Y_{ll}(\varOmega)
\end{align}
and 
\begin{align}
r^l\exp[-\nu\bvec{r}^2] \equiv \frac{1}{t_l}R_{0l}(r),
\label{Eq.RadialWaveFunction}
\end{align}
and $s_l$ and $t_l$ are the normalization factors of $Y_{ll}(\varOmega)$ and $R_{0l}(r)$, respectively.
The proton $i=1$ has spin-up, and the spherical harmonics $Y_{ll}(\varOmega)$ with spin-up has $j_z=l+1/2$, which only couples to $j=l+1/2$ (stretched configuration), and the spin-orbit interaction works attractively.
Thus the proton $i=1$ is described as a linear combination of $j$-upper orbits with $j=l+1/2$ and $j_z = j$,
\begin{align}
\phi_{i=1}=\sum_{j=1/2}^\infty{}a_jR^{j-\frac{1}{2}} \langle \bvec{r} |j,j\rangle\tau_1,
\label{Eq.phi1ExpandedByjj}
\end{align}
where $a_j$ is a coefficient for the $\langle \bvec{r} |j,j\rangle$ orbit with a separated factor of $R^{j-\frac{1}{2}}$.
The proton $i=2$ is the time reversal partner of $i=1$ with spin-down,
\begin{align}
\phi_{i=2}=\sum_{j=1/2}^\infty{}a_{-j}R^{j-\frac{1}{2}} \langle \bvec{r} |j,-j\rangle\tau_2.
\label{Eq.phi2ExpandedByjj}
\end{align} 
\par
For other protons, $i=3$ and $4$ are introduced as in the same quasi cluster, and $i=5$ and $6$ also belong to the same quasi cluster, but this is different from the one for $i=3$ and $i=4$.
Their wave functions are introduced by rotating both the spatial and spin parts of the protons $i=1,2$ about the $y$ axis as
\begin{align}
\phi_{i+2}=\hat{R}(\alpha=0, \beta=\theta_1, \gamma=0)\phi_i, \\
\phi_{i+4}=\hat{R}(\alpha=0, \beta=\theta_2, \gamma=0)\phi_i,
\end{align}
where $i=1,2$.
The rotation does not change the total angular momentum $j$, and the resultant single-particle wave functions are also linear combinations of $j$-upper orbits.
Here, $\alpha$, $\beta$, $\gamma$ are the Euler angles, and $\hat{R}$($\alpha$, $\beta$, $\gamma$) is the rotation operator.
The parameters $\theta_1$ and $\theta_2$ are rotational angles, and they are introduced as $\theta_1=2\pi/3$ and $\theta_2=4\pi/3$, which give an equilateral triangular shape of the three $\alpha$ clusters when $\varLambda$ is equal to zero.
The $\langle \bvec{r}|j,j\rangle$ orbit after the rotation can be expressed as
\begin{align}
\hat{R}(\alpha=0, \beta=\theta, \gamma=0)\langle \bvec{r}|j,j\rangle=\sum_{m=-j}^jd^j_{mj}(\theta)\langle \bvec{r}|j,m\rangle,
\end{align}
where $d^j_{km}(\beta)$ is Wigner's small $d$ function,
\begin{align}
\langle{}j,k|\hat{R}(\alpha,\beta,\gamma)|j,m\rangle=\exp[-ik\alpha]d^j_{km}(\beta)\exp[-im\gamma].
\end{align}
Thus the rotated single-particle wave function is expressed as
\begin{align}
\hat{R}(\alpha=0, \beta=\theta, \gamma=0)\phi_1=\sum_{j=1/2}^\infty\sum_{m=-j}^ja_jR^{j-\frac{1}{2}}d^j_{mj}(\theta)
\langle \bvec{r}|j,m\rangle\tau_1.
\end{align}
The result shows that when $\varLambda$ is equal to unity, all the single-particle wave functions are described as the linear combinations of $j$-upper orbits, and the Slater determinant has only the $(0s_{1/2})^4(0p_{3/2})^8$ component at the lowest order of $R$.

\subsection{Extension of AQCM}
\label{Sec.TheModel}
Here we explain our new model, which is the extension of AQCM.

\subsubsection{Total wave function}
The total wave function is expressed as a linear combination of different Slater determinants based on the generator coordinate method (GCM) as
\begin{align}
\varPhi^n=&\sum_kc_k^n\varPsi_k,
\\
\varPsi_k=&\hat{P}^J_{MK}\hat{P}^\pi\mathcal{A}[\phi_1\cdots\phi_{12}]_k,
\end{align}
where $\hat{P}^J_{MK}$ and $\hat{P}^\pi$ are the angular momentum and parity projection operators.
Here $k=1,2\ldots$ is a label for different basis states.
The coefficients $\{ c_k^n \}$ are determined by solving the Hill-Wheeler equation, and $n=1,2,\ldots$ denotes the $n$-th excited state obtained after the diagonalization of the Hamiltonian.
In this paper, we particularly pay attention to the $0^+$ states, thus $J=M=K=0$ and $\pi=+$.  \par
\begin{figure}
\centering
\includegraphics[width=.4\textwidth, trim= 0cm 0cm 0cm 0cm, clip]{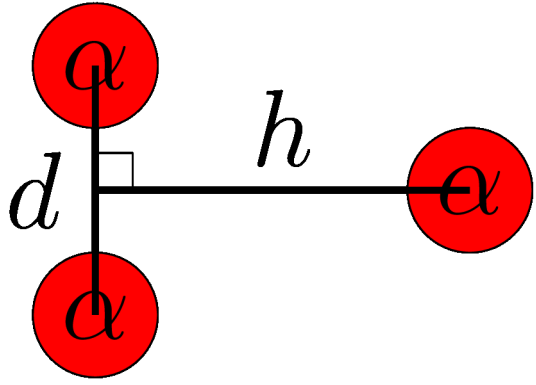}
\caption{(Color online) Schematic figure for the cluster model configurations.
The red spheres show the $\alpha$ clusters.}
\label{Fig.model}
\end{figure}
For the basis states, we prepare both the shell and cluster model ones.
For the shell model part, we use AQCM, and in addition to the subclosure configuration of $0p_{3/2}$, we introduce five different two particle two hole (2p2h) configurations.
For the cluster model space, we introduce thirty different three $\alpha$ configurations.
In total, we superpose $6+30=36$ basis states and diagonalize the Hamiltonian. 
For the width parameter $\nu$ $(=1/2b^2)$ in Eq.~(\ref{Eq.SpatialPart}), we take $b=1.4\,\mathrm{fm}$. \par
For the shell model basis states, as shown in the previous subsection, we can transform the $\alpha$ cluster model wave function to the $0p_{3/2}$ subclosure configuration of the $jj$-coupling shell model using AQCM.
This is $(0s_{1/2})^2(0p_{3/2})^4$ for the proton and neutron parts, and here we call it zero particle zero hole (0p0h) state.
To generate this state, we take a small enough $R$ value of $R=0.1\,\mathrm{fm}$ in Eqs.~(\ref{Eq.zeta1}) and (\ref{Eq.zeta2}).
In addition, we newly introduce five 2p2h configurations.
Four of them correspond to the normal BCS-like pairing effect of protons or neutrons; $(0s_{1/2})^2(0p_{3/2})^2(0p_{1/2})^2$ and $(0s_{1/2})^2(0p_{3/2})^2(0d_{5/2})^2$ are introduced for the proton part or neutron part.
We further taken into account the proton-neutron pairing effect.
For this purpose, we prepare a basis state, where one proton and one neutron are excited from $0p_{3/2}$ to $0p_{1/2}$; $(0s_{1/2})^2(0p_{3/2})^3(0p_{1/2})^1$ for both proton and neutron parts.
This is also 2p2h, but each isospin (proton or neutron) part is one particle one hole (1p1h).
In total we introduce six configurations for the $0^+$ states of $^{12}\mathrm{C}$; 0p0h for both proton and neutron parts ($pn$-0p0h), 2p2h excitation to $0p_{1/2}$ for the proton part ($pp$-$p_{1/2}$-2p2h), that for the neutron part ($nn$-$p_{1/2}$-2p2h), 2p2h excitation to $0d_{5/2}$ for the proton part ($pp$-$d_{5/2}$-2p2h), that for the neutron part ($nn$-$d_{5/2}$-2p2h), and 1p1h to $0p_{1/2}$ for both proton and neutron parts ($pn$-$p_{1/2}$-2p2h).
The 1p1h configuration is explained in Sec.~\ref{Sec.DescriptionOfOneParticleOneHole},
and the 2p2h configurations are explained in Sec.~\ref{Sec.DescriptionOfTwoParticleTwoHoleByAntisymmetrizationEffect} and
Sec.~\ref{Sec.DescriptionOf0s1/20p3/20d5/2ConfigurationByARegularTriangleStructure}.
\par
For the cluster model basis states, as schematically shown in Fig.~\ref{Fig.model}, the configurations are introduced with isosceles triangular shapes.
The parameters $d$ and $h$ are the base and height of the isosceles triangle, respectively, and they are taken as $d=1,2,\ldots,5\,\mathrm{fm}$ and $h=1,2,\ldots,6\,\mathrm{fm}$.
There are $5\times6=30$ basis states for the cluster model side.

\subsubsection{Hamiltonian}
The Hamiltonian used in the present calculation is
\begin{align}
\hat{H}=\hat{T}-\hat{T}_\mathrm{G}+\hat{V}_\mathrm{C}+\hat{V}_\mathrm{LS}+\hat{V}_\mathrm{Coulomb},
\end{align}
where $\hat{T}$ is the total kinetic energy operator and $\hat{T}_\mathrm{G}$ is the kinetic energy operator of the center of mass motion.
For the central force $\hat{V}_\mathrm{C}$, we use the Volkov No.2 force~\cite{Volkov.NuclPhys74.33.1965} given by
\begin{align}
\hat{V}_\mathrm{C}=\sum_{i<j}^A\left[V_a\exp\left(-\frac{\hat{\bvec{r}}_{ij}^2}{\alpha^2}\right)+V_r\exp\left(-\frac{\hat{\bvec{r}}_{ij}^2}{\rho^2}\right)\right]\left[W+B\hat{P}^\sigma_{ij}-H\hat{P}^\tau_{ij}-M\hat{P}^\sigma_{ij}\hat{P}^\tau_{ij}\right],
\end{align}
where $V_a=-60.65\,\mathrm{MeV}$, $V_r=61.14\,\mathrm{MeV}$, $\alpha=1.80\,\mathrm{fm}$, and $\rho=1.01\,\mathrm{fm}$ are the original values.
We take $M=1-W=0.6$.
Here, $B$ and $H$ denote the Bartlett and Heisenberg terms, which are added to remove the bound state of two neutrons.
We take $B=H=0.125$.
For the spin-orbit force $\hat{V}_\mathrm{LS}$, we use the spin-orbit part of the G3RS force~\cite{Tamagaki.ProgTheorPhys39.91.1968} given by
\begin{align}
\hat{V}_\mathrm{LS}=\sum_{i<j}^A\left[V_\mathrm{LS1}\exp\left(-\frac{\hat{\bvec{r}}_{ij}^2}{\eta_1^2}\right)+V_\mathrm{LS2}\exp\left(-\frac{\hat{\bvec{r}}_{ij}^2}{\eta_2^2}\right)\right]\hat{P}_{ij}(^3O)\hat{\bvec{L}}_{ij}\cdot\hat{\bvec{S}}_{ij},
\end{align}
where $\eta_1=0.447\,\mathrm{fm}$ and $\eta_2=0.6\,\mathrm{fm}$ are the original values.
The coefficients $V_\mathrm{LS1}=-V_\mathrm{LS2}=1600\,\mathrm{MeV}$ are determined to give a reasonable energy for the ground state in $^{12}\mathrm{C}$.
Also, the validity of $V_\mathrm{LS1}=-V_\mathrm{LS2}=1600\,\mathrm{MeV}$ is checked in Ref~\cite{Suhara.PhysRevC87.054334.2013}.
The operator $\hat{V}_\mathrm{Coulomb}$ is the Coulomb potential for protons.

\section{Results and discussion}
\label{Sec.ResultsAndDiscussion}
As already seen, the closure configurations of the major shells can be described by conventional $\alpha$ cluster models, and subclosure configurations of the $jj$-coupling shell model can be described by AQCM.
Here we extend AQCM.
At first we discuss the AQCM wave functions with general $\varLambda$ values and next show how to describe particle hole excitations.
For $^{12}\mathrm{C}$, single particle excitations from $0p_{3/2}$ to $0p_{1/2}$ and $0d_{5/2}$ are introduced for the proton part, neutron part, and proton-neutron part, and the effect of BCS-like pairing is incorporated.
Finally these shell-model-like wave functions are coupled with the three $\alpha$ cluster wave functions.

\subsection{AQCM wave functions with general $\varLambda$ values}
We already discussed that $\varLambda = 0$ corresponds to $\alpha$ cluster states and $\varLambda=1$ with small $R$ corresponds to the $jj$-coupling shell model states.
However, the discussion for the general $\varLambda$ values ($\varLambda\neq 0,1$) is insufficient.
In this subsection, we investigate the feature of the AQCM wave functions with general $\varLambda$ values in $^{12}\mathrm{C}$. \par
Using the relations for the spherical harmonics
\begin{align}
\frac{x+iy}{r}=&\frac{1}{s_1}Y_{11}(\varOmega),
\\
\frac{x-iy}{r}=&\frac{1}{s_{-1}}Y_{1-1}(\varOmega)=-\frac{1}{s_1}Y_{1-1}(\varOmega),
\end{align}
and Eqs.~(\ref{Taylor}) and (\ref{Eq.RadialWaveFunction}), the single-particle wave function of the proton $i=1$ 
[Eq.~(\ref{spwf1})] with general $\varLambda$ becomes
\begin{align}
\phi_{i=1}=&\left(\frac{2\nu}{\pi}\right)^{\frac{3}{4}}e^{-\nu{}R^2(1-\varLambda^2)} \nonumber \\
&\times\left[\frac{1}{s_0t_0}R_{00}(r)Y_{00}(\varOmega)+\frac{2\nu{}R}{s_1t_1}R_{01}(r)\left(\frac{1+\varLambda}{2}Y_{11}(\varOmega)-\frac{1-\varLambda}{2}Y_{1-1}(\varOmega)\right)+\mathcal{O}(R^2)\right]\chi_\uparrow\tau_1.
\end{align}
We introduce $jj$-coupling bases using the Clebsch-Gordan coefficients,
\begin{align}
\ket{j=\frac{3}{2},j_z=\frac{1}{2}}=&R_{01}(r)\left(\frac{1}{\sqrt{3}}Y_{11}(\varOmega)\chi_\downarrow+\sqrt{\frac{2}{3}}Y_{10}(\varOmega)\chi_\uparrow\right),
\label{Eq.3/21/2} \\
\ket{j=\frac{3}{2},j_z=-\frac{1}{2}}=&R_{01}(r)\left(\sqrt{\frac{2}{3}}Y_{10}(\varOmega)\chi_\downarrow+\frac{1}{\sqrt{3}}Y_{1-1}(\varOmega)\chi_\uparrow\right),
\label{Eq.3/2-1/2} \\
\ket{j=\frac{1}{2},j_z=\frac{1}{2}}=&R_{01}(r)\left(\sqrt{\frac{2}{3}}Y_{11}(\varOmega)\chi_\downarrow-\frac{1}{\sqrt{3}}Y_{10}(\varOmega)\chi_\uparrow\right),
\label{Eq.1/21/2} \\
\ket{j=\frac{1}{2},j_z=-\frac{1}{2}}=&R_{01}(r)\left(\frac{1}{\sqrt{3}}Y_{10}(\varOmega)\chi_\downarrow-\sqrt{\frac{2}{3}}Y_{1-1}(\varOmega)\chi_\uparrow\right).
\label{Eq.1/2-1/2}
\end{align}
Thus the single-particle wave function of proton $i=1$ with general $\varLambda$ becomes
\begin{align}
\phi_{i=1}=&e^{-\nu{}R^2(1-\varLambda^2)} \nonumber \\
&\times\left[a_{1/2}\ket{s\frac{1}{2}\frac{1}{2}}\right. \nonumber \\
&\hspace{3ex} \left.+a_{3/2}R\left(\frac{1+\varLambda}{2}\ket{p\frac{3}{2}\frac{3}{2}}-\frac{1-\varLambda}{2}\left(\frac{1}{\sqrt{3}}\ket{p\frac{3}{2}\frac{-1}{2}}-\sqrt{\frac{2}{3}}\ket{p\frac{1}{2}\frac{-1}{2}}\right)\right)+\mathcal{O}(R^2)\right]\tau_1,
\end{align}
where $s$ and $p$ are indexes to distinguish the $s$ and $p$ orbits.
Similarly, the single-particle wave function of the proton $i=2$, which is time reversal of $i=1$, becomes
\begin{align}
\phi_{i=2}=&e^{-\nu{}R^2(1-\varLambda^2)} \nonumber \\
&\times\left[a_{-1/2}\ket{s\frac{1}{2}\frac{-1}{2}}\right. \nonumber \\
&\hspace{3ex} \left.+a_{-3/2}R\left(-\frac{1-\varLambda}{2}\left(\frac{1}{\sqrt{3}}\ket{p\frac{3}{2}\frac{1}{2}}+\sqrt{\frac{2}{3}}\ket{p\frac{1}{2}\frac{1}{2}}\right)+\frac{1+\varLambda}{2}\ket{p\frac{3}{2}\frac{-3}{2}}\right)+\mathcal{O}(R^2)\right]\tau_2.
\end{align}
The single-particle wave functions of protons $i=3-6$ are generated by multiplying the rotational operators $\hat{R}(0,2\pi/3,0)$ or $\hat{R}(0,4\pi/3,0)$ for the single-particle wave functions of protons $i=1,2$ as in original AQCM~\cite{Suhara.PhysRevC87.054334.2013},
\begin{align}
\phi_{i=3}=&e^{-\nu{}R^2(1-\varLambda^2)} \nonumber \\
&\times\left[a_{1/2}\left(\frac{1}{2}\ket{s\frac{1}{2}\frac{1}{2}}+\frac{\sqrt{3}}{2}\ket{s\frac{1}{2}\frac{-1}{2}}\right)\right. \nonumber \\
&\hspace{1eM}+a_{3/2}R\left(-\frac{1-2\varLambda}{8}\ket{p\frac{3}{2}\frac{3}{2}}+\frac{1+2\varLambda}{8}\ket{p\frac{3}{2}\frac{1}{2}}-\sqrt{2}\frac{1-\varLambda}{4}\ket{p\frac{1}{2}\frac{1}{2}}\right. \nonumber \\
&\left.\left.\hspace{2eM}+\frac{1}{\sqrt{3}}\frac{7+2\varLambda}{8}\ket{p\frac{3}{2}\frac{-1}{2}}+\sqrt{\frac{2}{3}}\frac{1-\varLambda}{4}\ket{p\frac{1}{2}\frac{-1}{2}}+\sqrt{3}\frac{1+2\varLambda}{8}\ket{p\frac{3}{2}\frac{-3}{2}}\right)+\mathcal{O}(R^2)\right]\tau_3, 
\end{align}
\begin{align}
\phi_{i=4}=&e^{-\nu{}R^2(1-\varLambda^2)} \nonumber \\
&\times\left[a_{-1/2}\left(-\frac{\sqrt{3}}{2}\ket{s\frac{1}{2}\frac{1}{2}}+\frac{1}{2}\ket{s\frac{1}{2}\frac{-1}{2}}\right)\right. \nonumber \\
&\hspace{1eM}+a_{-3/2}R\left(-\sqrt{3}\frac{1+2\varLambda}{8}\ket{p\frac{3}{2}\frac{3}{2}}+\frac{1}{\sqrt{3}}\frac{7+2\varLambda}{8}\ket{p\frac{3}{2}\frac{1}{2}}-\sqrt{\frac{2}{3}}\frac{1-\varLambda}{4}\ket{p\frac{1}{2}\frac{1}{2}}\right. \nonumber \\
&\left.\left.\hspace{2eM} -\frac{1+2\varLambda}{8}\ket{p\frac{3}{2}\frac{-1}{2}}-\sqrt{2}\frac{1-\varLambda}{4}\ket{p\frac{1}{2}\frac{-1}{2}}-\frac{1-2\varLambda}{8}\ket{p\frac{3}{2}\frac{-3}{2}}\right)+\mathcal{O}(R^2)\right]\tau_4, 
\end{align}
\begin{align}
\phi_{i=5}=&e^{-\nu{}R^2(1-\varLambda^2)} \nonumber \\
&\times\left[a_{1/2}\left(-\frac{1}{2}\ket{s\frac{1}{2}\frac{1}{2}}+\frac{\sqrt{3}}{2}\ket{s\frac{1}{2}\frac{-1}{2}}\right)\right. \nonumber \\
&\hspace{1eM}+a_{3/2}R\left(\frac{1-2\varLambda}{8}\ket{p\frac{3}{2}\frac{3}{2}}+\frac{1+2\varLambda}{8}\ket{p\frac{3}{2}\frac{1}{2}}-\sqrt{2}\frac{1-\varLambda}{4}\ket{p\frac{1}{2}\frac{1}{2}}\right. \nonumber \\
&\left.\left.\hspace{2eM}-\frac{1}{\sqrt{3}}\frac{7+2\varLambda}{8}\ket{p\frac{3}{2}\frac{-1}{2}}-\sqrt{\frac{2}{3}}\frac{1-\varLambda}{4}\ket{p\frac{1}{2}\frac{-1}{2}}+\sqrt{3}\frac{1+2\varLambda}{8}\ket{p\frac{3}{2}\frac{-3}{2}}\right)+\mathcal{O}(R^2)\right]\tau_5, 
\end{align}
\begin{align}
\phi_{i=6}=&e^{-\nu{}R^2(1-\varLambda^2)} \nonumber \\
&\times\left[a_{-1/2}\left(-\frac{\sqrt{3}}{2}\ket{s\frac{1}{2}\frac{1}{2}}-\frac{1}{2}\ket{s\frac{1}{2}\frac{-1}{2}}\right)\right. \nonumber \\
&\hspace{1eM}+a_{-3/2}R\left(-\sqrt{3}\frac{1+2\varLambda}{8}\ket{p\frac{3}{2}\frac{3}{2}}-\frac{1}{\sqrt{3}}\frac{7+2\varLambda}{8}\ket{p\frac{3}{2}\frac{1}{2}}+\sqrt{\frac{2}{3}}\frac{1-\varLambda}{4}\ket{p\frac{1}{2}\frac{1}{2}}\right. \nonumber \\
&\left.\left.\hspace{2eM} -\frac{1+2\varLambda}{8}\ket{p\frac{3}{2}\frac{-1}{2}}-\sqrt{2}\frac{1-\varLambda}{4}\ket{p\frac{1}{2}\frac{-1}{2}}+\frac{1-2\varLambda}{8}\ket{p\frac{3}{2}\frac{-3}{2}}\right)+\mathcal{O}(R^2)\right]\tau_6.
\end{align}
Thus the proton part of the wave function becomes
\begin{align}
\varPsi_p=&\mathcal{A}[\phi_1,\ldots,\phi_6] \nonumber \\
=&-\frac{9}{16}e^{-6\nu{}R^2(1-\varLambda^2)}(a_{1/2}a_{-3/2}-a_{-1/2}a_{3/2})^2a_{3/2}a_{-3/2}R^4 \nonumber \\
&\times\mathcal{A}\left[\ket{s\frac{1}{2}\frac{1}{2}},\ket{s\frac{1}{2}\frac{-1}{2}},
-\frac{1}{2}\ket{p\frac{3}{2}\frac{3}{2}}+\frac{\sqrt{3}}{2}\ket{p\frac{3}{2}\frac{-1}{2}},
\frac{\sqrt{3}}{2}\ket{p\frac{3}{2}\frac{1}{2}}-\frac{1}{2}\ket{p\frac{3}{2}\frac{-3}{2}},\right. \nonumber \\
&\hspace{1eM} \left.-\frac{1-\varLambda}{\sqrt{2}}\ket{p\frac{1}{2}\frac{1}{2}}+\frac{1+2\varLambda}{\sqrt{3}}\ket{p\frac{3}{2}\frac{-3}{2}},
-\frac{1+2\varLambda}{\sqrt{3}}\ket{p\frac{3}{2}\frac{3}{2}}-\frac{1-\varLambda}{\sqrt{2}}\ket{p\frac{1}{2}\frac{-1}{2}}\right] \nonumber \\
&+\mathcal{O}(R^5),
\label{Eq.PsiWithOnlyLambda}
\end{align}
where we omit the isospin part of the wave function.
The neutron part is introduced in the completely same way.
In Eq.~(\ref{Eq.PsiWithOnlyLambda}), two protons occupy $0s$ orbits and the others are described by the superposition of four different $0p$ orbits.
We can easily check that $\varLambda=1$ gives the $(0s_{1/2})^2(0p_{3/2})^4$ configuration with a small enough $R$ value as in original AQCM~\cite{Suhara.PhysRevC87.054334.2013}.

\subsection{Description of 2p2h ($p_{1/2}$-2p2h) }
\label{Sec.DescriptionOfTwoParticleTwoHoleByAntisymmetrizationEffect}
We show that the AQCM wave functions describe the 2p2h excitations of protons from the subclosure configuration to $0p_{1/2}$ [$(0s_{1/2})^4(0p_{3/2})^6(0p_{1/2})^2$] by using a different $\varLambda$ parameter for each quasi-$\alpha$ cluster (completely the same procedure can be applied also to the neutron part).
We introduce two $\varLambda$ parameters, $\varLambda_a$ for protons $i=1-2$ and $\varLambda_b$ for protons $i=3-6$.
After the antisymmetrization, the proton part of the wave function becomes
\begin{align}
\varPsi_p=&\mathcal{A}[\phi_1,\ldots,\phi_6] \nonumber \\
=&e^{-2\nu{}R^2(3-\varLambda_a^2-2\varLambda_b^2)}a_{1/2}a_{-1/2}a_{3/2}^2a_{-3/2}^2R^4 \nonumber \\
&\times\mathcal{A}\left[\ket{s\frac{1}{2}\frac{1}{2}},\ket{s\frac{1}{2}\frac{-1}{2}}, \right. \nonumber \\
&\hspace{1eM} -\frac{3+2(\varLambda_a-\varLambda_b)}{8}\ket{p\frac{3}{2}\frac{3}{2}}+\frac{1}{\sqrt{3}}\frac{9-2(\varLambda_a-\varLambda_b)}{8}\ket{p\frac{3}{2}\frac{-1}{2}}+\sqrt{\frac{2}{3}}\frac{\varLambda_a-\varLambda_b}{4}\ket{p\frac{1}{2}\frac{-1}{2}}, \nonumber \\
&\hspace{1eM} \frac{1}{\sqrt{3}}\frac{9-2(\varLambda_a-\varLambda_b)}{8}\ket{p\frac{3}{2}\frac{1}{2}}-\sqrt{\frac{2}{3}}\frac{\varLambda_a-\varLambda_b}{4}\ket{p\frac{1}{2}\frac{1}{2}}-\frac{3+2(\varLambda_a-\varLambda_b)}{8}\ket{p\frac{3}{2}\frac{-3}{2}}, \nonumber \\
&\hspace{1eM} -\frac{1-2(\varLambda_a+\varLambda_b)}{4}\ket{p\frac{3}{2}\frac{1}{2}}-\frac{2-(\varLambda_a+\varLambda_b)}{\sqrt{2}}\ket{p\frac{1}{2}\frac{1}{2}}+\sqrt{3}\frac{3+2(\varLambda_a+\varLambda_b)}{4}\ket{p\frac{3}{2}\frac{-3}{2}}, \nonumber \\
&\hspace{1eM} \left.-\sqrt{3}\frac{3+2(\varLambda_a+\varLambda_b)}{4}\ket{p\frac{3}{2}\frac{3}{2}}+\frac{1-2(\varLambda_a+\varLambda_b)}{4}\ket{p\frac{3}{2}\frac{-1}{2}}-\frac{2-(\varLambda_a+\varLambda_b)}{\sqrt{2}}\ket{p\frac{1}{2}\frac{-1}{2}}\right] \nonumber \\
&+\mathcal{O}(R^5).
\label{Eq.PsipLambdaaLambdab}
\end{align}
Here, we used the time reversal relations, $a_{-1/2}=a_{1/2}$ and $a_{-3/2}=-a_{3/2}$, and omitted the isospin part.
As easily recognized, the conditions $3+2(\varLambda_a-\varLambda_b)=0$ and $3+2(\varLambda_a+\varLambda_b)=0$, namely $(\varLambda_a,\varLambda_b)=(-3/2,0)$, allow us to remove the components of $|p,3/2,3/2\rangle$ and $|p,3/2,-3/2\rangle$.
In this case, the proton part of the wave function [Eq.~(\ref{Eq.PsipLambdaaLambdab})] becomes
\begin{align}
\varPsi_p=&6e^{-\frac{3}{2}\nu{}R^2}a_{1/2}a_{-1/2}a_{3/2}^2a_{-3/2}^2R^4 \nonumber \\
&\times\mathcal{A}\left[\ket{s\frac{1}{2}\frac{1}{2}},\ket{s\frac{1}{2}\frac{-1}{2}},\ket{p\frac{3}{2}\frac{1}{2}},\ket{p\frac{3}{2}\frac{-1}{2}},\ket{p\frac{1}{2}\frac{1}{2}},\ket{p\frac{1}{2}\frac{-1}{2}}\right]+\mathcal{O}(R^5),
\end{align}
and it coincides with the $(0s_{1/2})^2(0p_{3/2})^2(0p_{1/2})^2$ configuration ($p_{1/2}$-2p2h) at the limit of $R\to0$.
By imposing other conditions, we can change the hole configuration.
The conditions $9-2(\varLambda_a-\varLambda_b)=0$ and $1-2(\varLambda_a+\varLambda_b)=0$, namely $(\varLambda_a,\varLambda_b)=(5/2,-2)$, allow us to remove the components of $|p,3/2,1/2\rangle$ and $|p,3/2,-1/2\rangle$.
Thus the proton part of the wave function [Eq.~(\ref{Eq.PsipLambdaaLambdab})] becomes
\begin{align}
\varPsi_p=&\frac{27}{8}e^{\frac{25}{2}\nu{}R^2}a_{1/2}a_{-1/2}a_{3/2}^2a_{-3/2}^2R^4 \nonumber \\
&\times\mathcal{A}\left[\ket{s\frac{1}{2}\frac{1}{2}},\ket{s\frac{1}{2}\frac{-1}{2}},\ket{p\frac{3}{2}\frac{3}{2}},\ket{p\frac{3}{2}\frac{-3}{2}},\ket{p\frac{1}{2}\frac{1}{2}},\ket{p\frac{1}{2}\frac{-1}{2}}\right]+\mathcal{O}(R^5),
\end{align}
and it coincides with the $(0s_{1/2})^2(0p_{3/2})^2(0p_{1/2})^2$ configuration ($p_{1/2}$-2p2h) at the limit of $R\to0$.
This is also the 2p2h excitation from $0p_{3/2}$ to $0p_{1/2}$, but the configuration is slightly different from the previous case.

\subsection{Description of 1p1h}
\label{Sec.DescriptionOfOneParticleOneHole}
Next we further improve the AQCM wave function to describe the 1p1h excitations from the subclosure configuration. 
For this purpose, we generalize the Gaussian center parameter in Eq.~(\ref{Eq.SpatialPart}) as
\begin{align}
\bvec{\zeta}_i=R(a_i\bvec{e}_x+ib_i\bvec{e}_y+c_i\bvec{e}_z),
\label{Eq.zeta.eq.aibici}
\end{align}
where $R$ is a real number with a dimension of length, and $a_i$, $b_i$, and $c_i$ are dimensionless real numbers.
Here $\bvec{e}_x$, $\bvec{e}_y$, and $\bvec{e}_z$ are unit vectors for the $x$, $y$, and $z$ axes, respectively.
The spin orientation is no longer fixed along the $z$ axis, and the spin wave function $\chi_i$ is more generalized as 
\begin{align}
\chi_i=\cos\frac{\beta_i}{2}\chi_\uparrow+\sin\frac{\beta_i}{2}\chi_\downarrow,
\label{Eq.chi.eq.cosbetaplussinbeta}
\end{align}
where $\beta_i$ is taken as a real parameter for simplicity.
If $(a_i,b_i,c_i)=(1,\varLambda,0)$ and $\beta_i=0$ are satisfied, Eq.~(\ref{Eq.zeta.eq.aibici}) coincides with original AQCM in Eq.~(\ref{Eq.zeta1}). \par
The single-particle wave function is expanded with the $jj$-coupling shell model bases, $\{|j,j_z\rangle\}$, as in the previous subsections.
Using the relation
\begin{align}
\frac{a_ix+ib_iy+c_iz}{r}=&A_i\frac{x+iy}{r}+B_i\frac{x-iy}{r}+C_i\frac{\sqrt{2}z}{r} \nonumber \\
=&\frac{1}{s_1}\left(A_iY_{11}(\varOmega)-B_iY_{1-1}(\varOmega)-C_iY_{10}(\varOmega)\right),
\end{align}
and Eqs.~(\ref{Eq.RadialWaveFunction}), (\ref{Eq.3/21/2})$-$(\ref{Eq.1/2-1/2}), this generalized single-particle wave function becomes
\begin{align}
\phi_i=&e^{-\nu{}R^2(a_i^2-b_i^2+c_i^2)} \nonumber \\
&\times\left[a_{1/2}\left(\cos\frac{\beta_i}{2}\ket{s\frac{1}{2}\frac{1}{2}}+\sin\frac{\beta_i}{2}\ket{s\frac{1}{2}\frac{-1}{2}}\right)\right. \nonumber \\
&\hspace{1eM}+a_{3/2}R\left(A_i\cos\frac{\beta_i}{2}\ket{p\frac{3}{2}\frac{3}{2}}+\frac{1}{\sqrt{3}}\left(A_i\sin\frac{\beta_i}{2}-\sqrt{2}C_i\cos\frac{\beta_i}{2}\right)\ket{p\frac{3}{2}\frac{1}{2}}\right. \nonumber \\
&\hspace{3eM}+\frac{1}{\sqrt{3}}\left(\sqrt{2}A_i\sin\frac{\beta_i}{2}+C_i\cos\frac{\beta_i}{2}\right)\ket{p\frac{1}{2}\frac{1}{2}}-\frac{1}{\sqrt{3}}\left(\sqrt{2}C_i\sin\frac{\beta_i}{2}+B_i\cos\frac{\beta_i}{2}\right)\ket{p\frac{3}{2}\frac{-1}{2}} \nonumber \\
&\hspace{3eM}\left.\left.-\frac{1}{\sqrt{3}}\left(C_i\sin\frac{\beta_i}{2}-\sqrt{2}B_i\cos\frac{\beta_i}{2}\right)\ket{p\frac{1}{2}\frac{-1}{2}}-B_i\sin\frac{\beta_i}{2}\ket{p\frac{3}{2}\frac{-3}{2}}\right)+\mathcal{O}(R^2)\right]\tau_i,
\label{Eq.GeneralAQCMSingleparticleWaveFunction}
\end{align}
where $A_i=(a_i+b_i)/2$, $B_i=(a_i-b_i)/2$, and $C_i=c_i/\sqrt{2}$, respectively.
The single-particle wave function has all the four components of $0p_{3/2}$ orbits and all the two components of $0p_{1/2}$ orbits with different coefficients. \par
Now a proton has all the six components of $0p$ orbits, and we remove some of them by imposing conditions.
If $A_i\sin(\beta_i/2)-\sqrt{2}C_i\cos(\beta_i/2)=0$ is satisfied, we can eliminate the component of $|p,3/2,1/2\rangle$.
Similarly, if $C_i\sin(\beta_i/2)-\sqrt{2}B_i\cos(\beta_i/2)=0$ is satisfied, the component of $|p,1/2,-1/2\rangle$ vanishes.
Thus, if $A_i\sin(\beta_i/2)-\sqrt{2}C_i\cos(\beta_i/2)=0$ and $C_i\sin(\beta_i/2)-\sqrt{2}B_i\cos(\beta_i/2)=0$ are simultaneously satisfied, the single-particle wave function does not have the $|p,3/2,1/2\rangle$ and $|p,1/2,-1/2\rangle$ components.
This was for one proton; however if all the protons satisfy the same conditions, the proton part of the wave function also does not have the components of $|p,3/2,1/2\rangle$ and $|p,1/2,-1/2\rangle$.
This is nothing but 1p1h excitation to $0p_{1/2}$. \par
\begin{table}[!h]
\caption{Example of
 $\{(a_i,ib_i,c_i)\}$ and coefficients for the spin wave functions for the six protons ($i=1-6$),  which describes the 1p1h configuration.
The required conditions are $a_i^2-b_i^2=2c_i^2$ and $\tan(\beta_i/2)=\mathrm{sign}(a_ic_i)\sqrt{2(a_i-b_i)/(a_i+b_i)}$. The center of mass of the system is set to the origin.
The parameters $a_i$, $b_i$, and $c_i$ are introduced in Eq.~(\ref{Eq.zeta.eq.aibici}), and the parameter $\beta_i$ is introduced in Eq.~(\ref{Eq.chi.eq.cosbetaplussinbeta}).
}
\label{Tab.ExampleOfTheSet}
\centering
\begin{tabular}{ccccccc}
\hline \hline
$i$ & 1 & 2 & 3 & 4 & 5 & 6 \\ \hline
$a_i$ & $1$ & $1$ & $-\frac{1}{2}$ & $-\frac{1}{2}$ & $-\frac{1}{2}$ & $-\frac{1}{2}$ \\
$ib_i$ & $i$ & $-i$ & $\frac{1}{2}i\varLambda$ & $-\frac{1}{2}i\varLambda$ & $\frac{1}{2}i\varLambda$ & $-\frac{1}{2}i\varLambda$ \\
$c_i$ & $0$ & $0$ & $\frac{\sqrt{1-\varLambda^2}}{2\sqrt{2}}$ & $\frac{\sqrt{1-\varLambda^2}}{2\sqrt{2}}$ & $-\frac{\sqrt{1-\varLambda^2}}{2\sqrt{2}}$ & $-\frac{\sqrt{1-\varLambda^2}}{2\sqrt{2}}$ \\
coefficient for $\chi_\uparrow$ ($\cos\frac{\beta_i}{2}$) & $1$ & $0$ & $\sqrt{\frac{1-\varLambda}{3+\varLambda}}$ & $\sqrt{\frac{1+\varLambda}{3-\varLambda}}$ & $\sqrt{\frac{1-\varLambda}{3+\varLambda}}$ & $\sqrt{\frac{1+\varLambda}{3-\varLambda}}$ \\
coefficient for $\chi_\downarrow$ ($\sin\frac{\beta_i}{2}$) & $0$ & $1$ & $-\sqrt{2}\sqrt{\frac{1+\varLambda}{3+\varLambda}}$ & $-\sqrt{2}\sqrt{\frac{1-\varLambda}{3-\varLambda}}$ & $\sqrt{2}\sqrt{\frac{1+\varLambda}{3+\varLambda}}$ & $\sqrt{2}\sqrt{\frac{1-\varLambda}{3-\varLambda}}$ \\ \hline \hline
\end{tabular}
\end{table}
In the following part, we simplify the conditions to describe 1p1h.
The conditions $A_i\sin(\beta_i/2)-\sqrt{2}C_i\cos(\beta_i/2)=0$ and $C_i\sin(\beta_i/2)-\sqrt{2}B_i\cos(\beta_i/2)=0$ are equivalent to $\tan(\beta_i/2)=\sqrt{2}C_i/A_i=\sqrt{2}B_i/C_i$.
Substituting $A_i=(a_i+b_i)/2$, $B_i=(a_i-b_i)/2$, and $C_i=c_i/\sqrt{2}$, the conditions become $a_i^2-b_i^2=2c_i^2$.
As $a_i$, $b_i$, and $c_i$ are real numbers, another condition of $|a_i|\ge|b_i|$ is required.
As a result, the conditions for the spin part of the wave function become $\tan(\beta_i/2)=\mathrm{sign}(a_ic_i)\sqrt{2(a_i-b_i)/(a_i+b_i)}$, where $\mathrm{sign}(\xi)=\xi/|\xi|$.
As an example which realizes the conditions, $a_i^2-b_i^2=2c_i^2$ and $\tan(\beta_i/2)=\mathrm{sign}(a_ic_i)\sqrt{2(a_i-b_i)/(a_i+b_i)}$, we show a set for the six protons in Table~\ref{Tab.ExampleOfTheSet}.
The center of mass of the system is set to the origin.
The parameters for protons $i=1$ and $2$ are equivalent to the ones for the original AQCM wave function~\cite{Suhara.PhysRevC87.054334.2013}.
We can confirm that the presence of a real parameter $\varLambda$ avoids the risk that some of the single particle orbits are not linear independent.
For the range of $\varLambda$, only  $0<\varLambda<1$ is allowable.
Using these parameters, the wave function of the proton part $\varPsi_p=\mathcal{A}[\phi_1,\ldots,\phi_6]$ describes the $(0s_{1/2})^2(0p_{3/2})^3(0p_{1/2})^1$ configuration at the limit of $R\to0$. \par
\begin{table}[!h]
\caption{Example of  $\{(a_i,ib_i,c_i)\}$ and coefficients for the spin wave functions for the six neutrons ($i=7-12$), which describes the 1p1h configuration.
The required conditions are $a_i^2-b_i^2=2c_i^2$ and $\tan(\beta_i/2)=-\mathrm{sign}(a_ic_i)\sqrt{(a_i-b_i)/[2(a_i+b_i)]}$. The center of mass of the system is set to the origin.
The parameters $a_i$, $b_i$, and $c_i$ are introduced in Eq.~(\ref{Eq.zeta.eq.aibici}), and the parameter $\beta_i$ is introduced in Eq.~(\ref{Eq.chi.eq.cosbetaplussinbeta}).
}
\label{Tab.ExampleOfTheSet2}
\centering
\begin{tabular}{ccccccc}
\hline \hline
$i$ & 7 & 8 & 9 & 10 & 11 & 12 \\ \hline
$a_i$ & $1$ & $1$ & $-\frac{1}{2}$ & $-\frac{1}{2}$ & $-\frac{1}{2}$ & $-\frac{1}{2}$ \\
$ib_i$ & $i$ & $-i$ & $\frac{1}{2}i\varLambda$ & $-\frac{1}{2}i\varLambda$ & $\frac{1}{2}i\varLambda$ & $-\frac{1}{2}i\varLambda$ \\
$c_i$ & $0$ & $0$ & $\frac{\sqrt{1-\varLambda^2}}{2\sqrt{2}}$ & $\frac{\sqrt{1-\varLambda^2}}{2\sqrt{2}}$ & $-\frac{\sqrt{1-\varLambda^2}}{2\sqrt{2}}$ & $-\frac{\sqrt{1-\varLambda^2}}{2\sqrt{2}}$ \\
coefficient for $\chi_\uparrow$ ($\cos\frac{\beta_i}{2}$) & $1$ & $0$ & $\sqrt{2}\sqrt{\frac{1-\varLambda}{3-\varLambda}}$ & $\sqrt{2}\sqrt{\frac{1+\varLambda}{3+\varLambda}}$ & $\sqrt{2}\sqrt{\frac{1-\varLambda}{3-\varLambda}}$ & $\sqrt{2}\sqrt{\frac{1+\varLambda}{3+\varLambda}}$ \\
coefficient for $\chi_\downarrow$ ($\sin\frac{\beta_i}{2}$) & $0$ & $1$ & $\sqrt{\frac{1+\varLambda}{3-\varLambda}}$ & $\sqrt{\frac{1-\varLambda}{3+\varLambda}}$ & $-\sqrt{\frac{1+\varLambda}{3-\varLambda}}$ & $-\sqrt{\frac{1-\varLambda}{3+\varLambda}}$ \\ \hline \hline
\end{tabular}
\end{table}
We need to couple all the nucleons to $0^+$, and 1p1h for the proton part and that for the neutron part must be introduced as time reversal partners.
For the protons, we choose the parameters in Table~\ref{Tab.ExampleOfTheSet}, and the wave function becomes
\begin{align}
\mathcal{A}\left[\ket{s\frac{1}{2}\frac{1}{2}},\ket{s\frac{1}{2}\frac{-1}{2}},\ket{p\frac{3}{2}\frac{3}{2}},\ket{p\frac{1}{2}\frac{1}{2}},\ket{p\frac{3}{2}\frac{-1}{2}},\ket{p\frac{3}{2}\frac{-3}{2}}\right].
\end{align}
Thus the neutron part must be introduced as
\begin{align}
\mathcal{A}\left[\ket{s\frac{1}{2}\frac{1}{2}},\ket{s\frac{1}{2}\frac{-1}{2}},\ket{p\frac{3}{2}\frac{3}{2}},\ket{p\frac{3}{2}\frac{1}{2}},\ket{p\frac{1}{2}\frac{-1}{2}},\ket{p\frac{3}{2}\frac{-3}{2}}\right].
\end{align}
The conditions which eliminate the components of $|p,3/2,-1/2\rangle$ and $|p,1/2,1/2\rangle$ in Eq.~(\ref{Eq.GeneralAQCMSingleparticleWaveFunction}) are $\sqrt{2}C_i\sin(\beta_i/2)+B_i\cos(\beta_i/2)=0$ and $\sqrt{2}A_i\sin(\beta_i/2)+C_i\cos(\beta_i/2)=0$.
These conditions are equivalent to $a_i^2-b_i^2=2c_i^2$ and $\tan(\beta_i/2)=-\mathrm{sign}(a_ic_i)\sqrt{(a_i-b_i)/[2(a_i+b_i)]}$.
As an example which satisfies these conditions, we show a set of $\{(a_i,ib_i,c_i)\}$ and coefficients for the spin wave functions for the six neutrons in Table~\ref{Tab.ExampleOfTheSet2}.
The center of mass of the system is set to the origin.
We choose these parameters in Table~\ref{Tab.ExampleOfTheSet2} for the neutron part.

\subsection{Description of $(0s_{1/2})^2(0p_{3/2})^2(0d_{5/2})^2$ configuration ($d_{5/2}$-2p2h)}
In this subsection, we describe the 2p2h excitations to a higher major shell.
In the conventional Brink model, the excitation was described by changing the spatial configuration of $\alpha$ clusters.
The excitation to the $sd$ shell, $(0s)^2(0p)^2(0d)^2$ for the protons, was described by assuming a configuration that the three $\alpha$ clusters are on a straight line.
Because of the antisymmetrization effect, two protons are excited from $0s$ to $0p$ orbits, and two protons are further excited to $0d$ orbits.
At that time there was no spin-orbit effect, but now we have to transform this $0d$ orbits to $0d_{5/2}$ orbits of the $jj$-coupling shell model.
Here we describe the $(0s_{1/2})^2(0p_{3/2})^2(0d_{5/2})^2$ configuration ($d_{5/2}$-2p2h) of $^{12}\mathrm{C}$ in two ways.
If we take up to the second order of $R$, the single-particle wave functions in Eqs.~(\ref{Eq.phi1ExpandedByjj}) and (\ref{Eq.phi2ExpandedByjj}) become
\begin{align}
\phi_{i=1}=&\left[a_{1/2}\ket{\frac{1}{2}\frac{1}{2}}+a_{3/2}R\ket{\frac{3}{2}\frac{3}{2}}+a_{5/2}R^2\ket{\frac{5}{2}\frac{5}{2}}+\mathcal{O}(R^3)\right]\tau_1, \\
\phi_{i=2}=&\left[a_{-1/2}\ket{\frac{1}{2}\frac{-1}{2}}+a_{-3/2}R\ket{\frac{3}{2}\frac{-3}{2}}+a_{-5/2}R^2\ket{\frac{5}{2}\frac{-5}{2}}+\mathcal{O}(R^3)\right]\tau_2,
\end{align}
respectively.
These two single-particle wave functions are used in both methods.

\subsubsection{Description of $(0s_{1/2})^2(0p_{3/2})^2(0d_{5/2})^2$ configuration by a linear structure}
Here we assume a linear shape, and the single-particle wave functions of protons $i=3,4$ are generated by multiplying the rotational operator $\hat{R}(\alpha=0,\beta=\pi,\gamma=0)$ for the protons $i=1,2$, respectively, and
\begin{align}
\phi_{i=3}=&\hat{R}(\alpha=0,\beta=\pi,\gamma=0)\phi_{i=1} \nonumber \\
=&\left[a_{1/2}\ket{\frac{1}{2}\frac{-1}{2}}+a_{3/2}R\ket{\frac{3}{2}\frac{-3}{2}}+a_{5/2}R^2\ket{\frac{5}{2}\frac{-5}{2}}+\mathcal{O}(R^3)\right]\tau_3, \\
\phi_{i=4}=&\hat{R}(\alpha=0,\beta=\pi,\gamma=0)\phi_{i=2} \nonumber \\
=&\left[-a_{-1/2}\ket{\frac{1}{2}\frac{1}{2}}-a_{-3/2}R\ket{\frac{3}{2}\frac{3}{2}}-a_{-5/2}R^2\ket{\frac{5}{2}\frac{5}{2}}+\mathcal{O}(R^3)\right]\tau_4.
\end{align}
Note that the rotation angle is $\pi$ so as to generate a linear structure.
The Gaussian center parameters of protons $i=5,6$ are set to the origin, and $\bvec{\zeta}_{i=5}=\bvec{\zeta}_{i=6}=\bvec{0}$, where $i=5$ and $6$ are spin-up and spin-down protons, respectively, 
\begin{align}
\phi_{i=5}=&a_{1/2}\ket{\frac{1}{2}\frac{1}{2}}\tau_5, \\
\phi_{i=6}=&a_{-1/2}\ket{\frac{1}{2}\frac{-1}{2}}\tau_6.
\end{align}
These six protons are arranged on a straight line, which creates additional nodes owing to the antisymmetrization effect.
Thus the proton part of the wave function becomes
\begin{align}
\varPsi_p=&\mathcal{A}[\phi_1,\ldots,\phi_6] \nonumber \\
=&-a_{1/2}a_{-1/2}(a_{3/2}a_{-5/2}-a_{-3/2}a_{5/2})^2R^6 \nonumber \\
&\times\mathcal{A}\left[\ket{\frac{1}{2}\frac{1}{2}},\ket{\frac{1}{2}\frac{-1}{2}},\ket{\frac{3}{2}\frac{3}{2}},\ket{\frac{3}{2}\frac{-3}{2}},\ket{\frac{5}{2}\frac{5}{2}},\ket{\frac{5}{2}\frac{-5}{2}}\right]+\mathcal{O}(R^7),
\end{align}
where the isospin part is omitted.
This wave function coincides with the $(0s_{1/2})^2(0p_{3/2})^2(0d_{5/2})^2$ configuration ($d_{5/2}$-2p2h) at the limit of $R\to0$.
The same procedure can be applied to the neutron part in the completely same way.

\subsubsection{Description of $(0s_{1/2})^2(0p_{3/2})^2(0d_{5/2})^2$ configuration by a regular triangle structure}
\label{Sec.DescriptionOf0s1/20p3/20d5/2ConfigurationByARegularTriangleStructure}
Here we describe the $(0s_{1/2})^2(0p_{3/2})^2(0d_{5/2})^2$ configuration ($d_{5/2}$-2p2h) in a different way; we do not assume a linear shape and the configuration remains with a regular triangular shape.
However, in return, the protons $i=3-6$, are rotated not about the $y$ axis but about the $z$ axis.
Then, not only the total angular momentum $j$, the $z$ component $j_z$ is unchanged after the rotation.
The single-particle wave functions of protons $i=3,4$ are generated by multiplying the rotational operator $\hat{R}(\alpha=2\pi/3,\beta=0,\gamma=0)$ to the protons $i=1,2$, respectively, and
\begin{align}
\phi_{i=3}=&\hat{R}(\alpha=2\pi/3,\beta=0,\gamma=0)\phi_{i=1} \nonumber \\
=&\left[a_{1/2}e^{-i\frac{\pi}{3}}\ket{\frac{1}{2}\frac{1}{2}}+a_{3/2}e^{-i\pi}R\ket{\frac{3}{2}\frac{3}{2}}+a_{5/2}e^{-i\frac{5\pi}{3}}R^2\ket{\frac{5}{2}\frac{5}{2}}+\mathcal{O}(R^3)\right]\tau_3, \\
\phi_{i=4}=&\hat{R}(\alpha=2\pi/3,\beta=0,\gamma=0)\phi_{i=2} \nonumber \\
=&\left[a_{-1/2}e^{i\frac{\pi}{3}}\ket{\frac{1}{2}\frac{-1}{2}}+a_{-3/2}e^{i\pi}R\ket{\frac{3}{2}\frac{-3}{2}}+a_{-5/2}e^{i\frac{5\pi}{3}}R^2\ket{\frac{5}{2}\frac{-5}{2}}+\mathcal{O}(R^3)\right]\tau_4.
\end{align}
The single-particle wave functions of protons $i=5,6$ are generated by multiplying the rotational operator $\hat{R}(\alpha=4\pi/3,\beta=0,\gamma=0)$ to the single-particle wave functions of protons $i=1,2$, respectively, and
\begin{align}
\phi_{i=5}=&\hat{R}(\alpha=4\pi/3,\beta=0,\gamma=0)\phi_{i=1} \nonumber \\
=&\left[a_{1/2}e^{-i\frac{2\pi}{3}}\ket{\frac{1}{2}\frac{1}{2}}+a_{3/2}e^{-2i\pi}R\ket{\frac{3}{2}\frac{3}{2}}+a_{5/2}e^{-i\frac{10\pi}{3}}R^2\ket{\frac{5}{2}\frac{5}{2}}+\mathcal{O}(R^3)\right]\tau_5, \\
\phi_{i=6}=&\hat{R}(\alpha=4\pi/3,\beta=0,\gamma=0)\phi_{i=2} \nonumber \\
=&\left[a_{-1/2}e^{i\frac{2\pi}{3}}\ket{\frac{1}{2}\frac{-1}{2}}+a_{-3/2}e^{2i\pi}R\ket{\frac{3}{2}\frac{-3}{2}}+a_{-5/2}e^{i\frac{10\pi}{3}}R^2\ket{\frac{5}{2}\frac{-5}{2}}+\mathcal{O}(R^3)\right]\tau_6.
\end{align}
These six protons are arranged in a regular triangular shape and the proton part of the wave function becomes
\begin{align}
\varPsi_p=&\mathcal{A}[\phi_1,\ldots,\phi_6] \nonumber \\
=&[(e^{-i\pi}-e^{-i\frac{\pi}{3}})(e^{-i\frac{10\pi}{3}}-e^{-i\frac{2\pi}{3}})-(e^{-2i\pi}-e^{-i\frac{2\pi}{3}})(e^{-i\frac{5\pi}{3}}-e^{-i\frac{\pi}{3}})] \nonumber \\
&\times[(e^{i\pi}-e^{i\frac{\pi}{3}})(e^{i\frac{10\pi}{3}}-e^{i\frac{2\pi}{3}})-(e^{2i\pi}-e^{i\frac{2\pi}{3}})(e^{i\frac{5\pi}{3}}-e^{i\frac{\pi}{3}})]a_{1/2}a_{-1/2}a_{3/2}a_{-3/2}a_{5/2}a_{-5/2}R^6 \nonumber \\
&\times\mathcal{A}\left[\ket{\frac{1}{2}\frac{1}{2}},\ket{\frac{1}{2}\frac{-1}{2}},\ket{\frac{3}{2}\frac{3}{2}},\ket{\frac{3}{2}\frac{-3}{2}},\ket{\frac{5}{2}\frac{5}{2}},\ket{\frac{5}{2}\frac{-5}{2}}\right]+\mathcal{O}(R^7),
\end{align}
where the isospin part is omitted.
This wave function coincides with the $(0s_{1/2})^2(0p_{3/2})^2(0d_{5/2})^2$ configuration ($d_{5/2}$-2p2h) at the limit of $R\to0$.
This is another method to create the $d_{5/2}$-2p2h configuration.
The same procedure can be applied to the neutron part.

\subsection{Energy levels and principal quantum numbers}
We couple all of the 2p2h configurations to the subclosure configuration of $0p_{3/2}$, and finally the three $\alpha$ cluster wave functions are mixed.
Concerning the $R$ and $\varLambda $ values, for the 1p1h configuration in Sec.~\ref{Sec.DescriptionOfOneParticleOneHole}, $R=0.1\,\mathrm{fm}$ and $\varLambda=0.1$ are employed, and for the 2p2h excitation to $0p_{1/2}$ in Sec.~\ref{Sec.DescriptionOfTwoParticleTwoHoleByAntisymmetrizationEffect}, we take $R=0.1\,\mathrm{fm}$ and $(\varLambda_a,\varLambda_b)=(-3/2,0)$.
For the 2p2h excitation to $0d_{5/2}$ in Sec.~\ref{Sec.DescriptionOf0s1/20p3/20d5/2ConfigurationByARegularTriangleStructure}, we take $R=0.1\,\mathrm{fm}$ in Eqs.~(\ref{Eq.zeta1}) and (\ref{Eq.zeta2}).
We discuss the obtained $0^+$ energy levels, principal quantum numbers, and $E0$ transition matrix elements.

\subsubsection{Energy levels with the shell model basis states}
\begin{figure}
\centering
\includegraphics[width=.6\textwidth, trim= 0cm 0cm 0cm 0cm, clip]{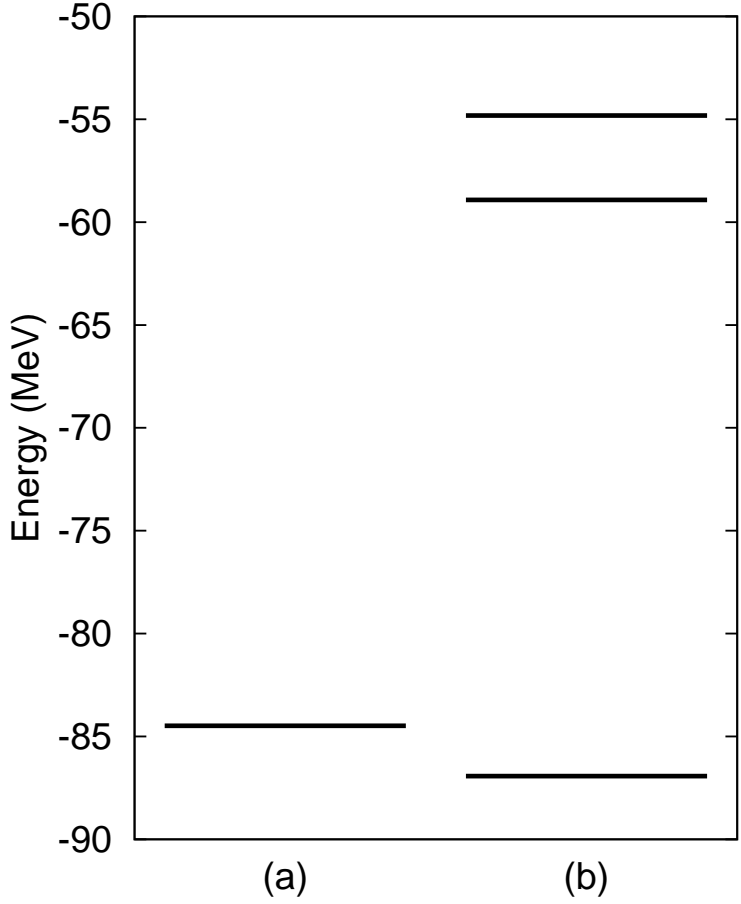}
\caption{$0^+$ energy levels of $^{12}\mathrm{C}$ calculated with shell model-like basis states.
 (a): $0^+$ energy of the subclosure configuration of $0p_{3/2}$ ($pn$-0p0h), (b):  $0^+$ levels obtained after coupling with the 2p2h states.}
\label{Fig.EnergyOfShell}
\end{figure}
We start with the shell model configurations introduced in Sec.~\ref{Sec.TheModel}.
Figure~\ref{Fig.EnergyOfShell} (a) shows the $0^+$ energy of $^{12}\mathrm{C}$ with the subclosure configuration of $0p_{3/2}$ ($pn$-0p0h), which is $-84.5\,\mathrm{MeV}$ (the experimental value $-92.2\,\mathrm{MeV}$ \cite{NNDC}).
In Fig.~\ref{Fig.EnergyOfShell} (b), the $0^+$ levels obtained after coupling with the 2p2h configurations are shown.
Here, we mixed five different 2p2h configurations; two nucleons are excited from $0p_{3/2}$ to $0p_{1/2}$ ($pp$-$p_{1/2}$-2p2h, $nn$-$p_{1/2}$-2p2h), or they are excited to $0d_{5/2}$ ($pp$-$d_{5/2}$-2p2h, $nn$-$d_{5/2}$-2p2h).
In addition, we couple a configuration that one proton and one neutron are excited from $0p_{3/2}$ to $0p_{1/2}$ ($pn$-$p_{1/2}$-2p2h).
In this way, we include the effects of BCS-like pairing for the proton part, neutron part, and proton-neutron part.
The energy of the ground state becomes $-86.9\,\mathrm{MeV}$, and this is lower than that of the subclosure configuration by $2.4\,\mathrm{MeV}$.
The reduction is caused by the coherent effects of the three BCS-like pairings.
The squared overlap between the ground state of the shell-model basis states and subclosure configuration of $0p_{3/2}$ ($pn$-0p0h) is $0.91$.

\subsubsection{Competition of shell and cluster structures}
\begin{table}[!h]
\caption{$0^+$ energies [$E\,(\mathrm{MeV})$] and principle quantum numbers ($N$) of $^{12}\mathrm{C}$ calculated using the shell (shell), cluster (cluster) model basis states.
The values for the mixed model space, subclosure configuration of $0p_{3/2}$ and cluster model basis states, are shown in the column ``$pn$-0p0h+cluster''.
The values for the full model space, shell and cluster basis states, are shown in the column ``shell+cluster''.}
\label{Tab.}
\centering
\begin{tabular}{cccccccccccc}
\hline \hline
 & \multicolumn{2}{c}{shell} & & \multicolumn{2}{c}{cluster} & & \multicolumn{2}{c}{$pn$-0p0h+cluster} & & \multicolumn{2}{c}{shell+cluster} \\ \cline{2-3}\cline{5-6}\cline{8-9}\cline{11-12}
 & $E$ & $N$ & & $E$ & $N$ & & $E$ & $N$ & & $E$ & $N$ \\ \hline
$0_1^+$ & $-86.9$ & $8.00$ & & $-89.1$ & $11.22$ & & $-91.8$ & $\ \, 9.40$ & & $-92.6$ & $\ \, 9.15$ \\
$0_2^+$ & $-58.9$ & $8.01$ & & $-79.1$ & $20.01$ & & $-83.2$ & $13.82$ & & $-83.4$ & $14.00$ \\ \hline \hline
\end{tabular}
\end{table}
Finally we couple the shell and cluster basis states.
In Table~\ref{Tab.}, the $0^+$ energies [$E\,(\mathrm{MeV})$] and principle quantum numbers ($N$) of $^{12}\mathrm{C}$ are shown.
The present interaction gives slightly lower ground state energy for the cluster basis states ($-89.1\,\mathrm{MeV}$) compared with the one for the shell model basis states ($-86.9\,\mathrm{MeV}$), but this is related to the fine tuning of the interaction parameters.
The ground state energy gets lower by $3.5\,\mathrm{MeV}$ from the one for the cluster model basis states by mixing both the shell and cluster model basis states ($-92.6\,\mathrm{MeV}$), since the spin-orbit interaction was not be taken into account within the cluster model basis states.
If we calculate without the 2p2h basis states, namely only within the subclosure configuration of $0p_{3/2}$ and cluster model basis states, the energy is $-91.8\,\mathrm{MeV}$.
This is higher by $0.8\,\mathrm{MeV}$ than the final result, and the mixing of the 2p2h configurations is found to have
a certain effect.
The principal quantum number for the ground state obtained with the shell model basis states is close to $8$, which is the lowest possible value, even though the 2p2h excitations to $0d_{5/2}$ are allowed.
On the other hand, the cluster model gives rather large value of $11.22$, and this is reduced to $9.15$ after coupling with the shell model basis states.
The three $\alpha$ configuration shrinks after coupling with the $jj$-coupling shell model states, as discussed in many preceding works including ours~\cite{Suhara2015,AMD-1,AMD-2,FMD-2}. \par
The $0_2^+$ state is the famous Hoyle state, which has the character of weakly interacting three $\alpha$ clusters.
Experimentally the state appears at $E_x = 7.65\,\mathrm{MeV}$, and our final result gives $9.2\,\mathrm{MeV}$.
Only within the cluster model basis states, the principal quantum number is $20.01$, and this is reduced to $14.00$ after coupling with the shell model basis states.
Since the ground state wave function is drastically changed after mixing the shell model basis states to the cluster configurations, the $0_2^+$ state is also influenced because of the orthogonal condition~\cite{Itagaki-NP}.
The matrix element of the $E0$ transition between the $0_1^+$ and $0_2^+$ states is $7.36\,e\,\mathrm{fm}^2$, which is $9.22\,e\,\mathrm{fm}^2$ only within the cluster model basis states (experimental value is $5.52\,e\,\mathrm{fm}^2$). \par
\begin{table}[!h]
\caption{Squared overlaps between the $0_{1,2}^+$ states and the six shell model basis states.}
\label{Tab.SquaredOverlap}
\centering
\begin{tabular}{ccccccc}
\hline \hline
 & $pn$-0p0h & $pn$-$p_{1/2}$-2p2h & $pp$-$p_{1/2}$-2p2h & $nn$-$p_{1/2}$-2p2h & $pp$-$d_{5/2}$-2p2h & $nn$-$d_{5/2}$-2p2h \\ \hline
$0_1^+$ & $4.21\times10^{-1}$ & $3.96\times10^{-2}$ & $6.78\times10^{-2}$ & $6.86\times10^{-2}$ & $9.26\times10^{-4}$ & $9.64\times10^{-4}$ \\
$0_2^+$ & $3.28\times10^{-1}$ & $1.10\times10^{-4}$ & $4.30\times10^{-4}$ & $5.27\times10^{-4}$ & $8.41\times10^{-4}$ & $7.92\times10^{-4}$ \\ \hline \hline
\end{tabular}
\end{table}
In Table~\ref{Tab.SquaredOverlap}, we show the squared overlaps between the $0_{1,2}^+$ states obtained with the full model space and the six shell model basis states introduced in the calculation.
The squared overlap between the $0_1^+$ state and the $pn$-0p0h is $0.42$, which is reduced from the one only within the shell model basis states ($0.91$).
This is because, using the present interaction parameters, the cluster model basis states gives lower ground state energy compared with the shell model basis states; however this tendency may change when we use slightly different parameter set.
Concerning the 2p2h configurations, the squared overlaps between the $0_1^+$ state and the $pn$-, $pp$-, and $nn$-$p_{1/2}$-2p2h are about $0.04-0.07$, which are not negligible.
However, the squared overlap between the $0_1^+$ state and $pp$-$d_{5/2}$-2p2h or $nn$-$d_{5/2}$-2p2h is more than an order of magnitude smaller.
For the $0_2^+$ state, it has the squared overlap with $pn$-0p0h by $0.33$, but the squared overlaps with the 2p2h configurations are quite small.

\section{Conclusion}
\label{Sec.Conclusion}
We have developed the framework of AQCM to describe not only the subclosure configuration of the $jj$-coupling shell model but also the 2p2h configurations, in addition to the cluster model wave functions.
In $^{12}\mathrm{C}$, it was shown that the 2p2h excitations from $0p_{3/2}$ to $0p_{1/2}$ and that to $0d_{5/2}$ were successfully described, which enables us to include the effects of BCS-like pairing for the proton part, neutron part, and proton-neutron part.
The correlation energy from the optimal configuration can be estimated not only in the cluster part but also in the shell model part. \par
For the ground $0^+$ state of $^{12}\mathrm{C}$, the interaction of the present calculation gives slightly lower energy for the cluster model basis states ($-89.1\,\mathrm{MeV}$) compared with the one for the shell model basis states ($-86.9\,\mathrm{MeV}$), and the ground state energy gets lower by $3.5\,\mathrm{MeV}$ by mixing both the shell and cluster model basis states ($-92.6\,\mathrm{MeV}$).
This is because the spin-orbit interaction is not be taken into account within the cluster model basis states.
If we calculate without the 2p2h basis states, the energy becomes $-91.8\,\mathrm{MeV}$, about $0.8\,\mathrm{MeV}$ higher, and the mixing of the 2p2h configurations is found to have a certain effect.
Only within the cluster model basis states, the principal quantum number is rather large, $11.22$, and this is reduced to $9.15$ after coupling with the shell model basis states. 
The three $\alpha$ configuration shrinks after coupling with the $jj$-coupling shell model states.
The squared overlap between the ground $0^+$ state and the 0p0h configuration of the $jj$-coupling shell model is $0.42$, and the overlaps with some of the 2p2h configurations are $0.04-0.07$, which are not negligible. \par
The $0_2^+$ state is the famous Hoyle state, and the present model gives $E_x = 9.2\,\mathrm{MeV}$.
The cluster model basis states give the principal quantum number of $20.01$, and this is reduced to $14.00$ after coupling with the shell model basis states.
Since the ground  state wave function is drastically changed after mixing the shell model basis states to the cluster configurations, the $0_2^+$ state is also influenced because of the orthogonal condition. \par
The method of describing particle hole excitations is considered to be applied to other light or even heavier nuclei.
As an example, description of four particle for hole configurations such as $(0s_{1/2})^4(0p_{3/2})^8(0d_{5/2})^4$ and the coupling with the cluster model wave functions ($^{12}\mathrm{C}$+$\alpha$, four $\alpha$'s) are going on for $^{16}\mathrm{O}$; the understanding of ``the mysterious $0^+$ state'' is a long-standing problem~\cite{Haxton}.
The systematic description of competition between particle hole excitations and cluster states is a challenging subject to be performed in near future.



\section*{Acknowledgments}
The authors are grateful for the computer facility at Yukawa Institute for Theoretical Physics, Kyoto University.
This work was supported by JSPS KAKENHI Grant Number 17K05440.

\end{document}